\documentstyle[prl,aps,multicol]{revtex}
\def\IM{\mathop{\Im m}\nolimits}
\def\RE{\mathop{\Re e}\nolimits}
\def\Z{{\bf Z}}

\begin{document}
\draft
\title{Charge fractionization 
in N=2 supersymmetric QCD}
\author{Frank Ferrari\footnote{email address: {\tt ferrari@physique.ens.fr}}}
\address{Laboratoire de Physique Th\'eorique
de l'\'Ecole
Normale Sup\'erieure\footnote{Unit\'e Propre de Recherche 701
du CNRS, associ\'ee
\`a l'\'Ecole Normale Sup\'erieure et \`a l'Universit\'e Paris Sud.} \\
24 rue Lhomond, 75231 Paris Cedex 05}
\date{September 1996}
\maketitle
\setbox1=\hbox{LPTENS-96/54}
\setbox2=\hbox{\tt hep-th@xxx/9609101}
\makeatletter
\global\@specialpagefalse
\def\@oddhead{\hfill\vbox{\box1\box2}}
\let\@evenhead\@oddhead
\begin{abstract}
It is shown that the physical ``quark number'' charges
which appear in the central charge of the supersymmetry algebra of $N=2$
supersymmetric QCD can take irrational values and depend non trivially
on the Higgs expectation value. This gives a physical interpretation of
the constant shifts which the ``electric'' and ``magnetic'' variables 
$a_D$ and $a$ undergo when encircling a singularity, and show
that duality in this model is truly an electric-magnetic-quark number
duality. Also included is a computation of the monodromy matrices
directly in the microscopic theory.
\end{abstract}
\begin{multicols}{2}
During the last couple of years, huge progress has been made in
the understanding of four dimensional
$N=2$ supersymmetric gauge theories, following the work of
Seiberg and Witten \cite{SWI,SWII}, where the low energy
(wilsonian) effective action was computed exactly up to two
derivatives and four fermions terms for the gauge group SU(2). 
The mathematical structure that emerged there was then generalized
to obtain very plausible solutions for
general gauge groups $G$ (see \cite{Ler} and \cite{AAG} with
references therein).
All these theories have a complex manifold of inequivalent vacua (moduli
space). This
degeneracy comes from flat directions in the scalar potential which
cannot be lifted quantum mechanically due to tight constraints
imposed by $N=2$ supersymmetry \cite{Seiberg}. The moduli space
has a Coulomb branch where the gauge group,
of rank $r$, is spontaneously broken
down to $U(1)^r$.
Many interesting physical phenomena occur along the Coulomb branch,
including the appearance of
massless solitonic states at strong coupling \cite{SWI,SWII}, 
a subtle realization
of electric-magnetic duality at the level of the low energy physics
\cite{SWI,SWII}, the existence of non-trivial Conformal
Field Theory in 
four dimensions \cite{Arg}
and discontinuities in the spectrum of stable BPS states
\cite{FB}. These ``exactly soluble'' 
theories are likely to play an outstanding
r\^ ole in our understanding of more
realistic gauge theories as QCD, like the Ising model
did for critical phenomena.

In this letter is explained how to compute from first principles
some abelian charges appearing in the central charge $Z$ of the
supersymmetry algebra. 
Then, knowing $Z$, one can compute the mass $m$ of any state
lying in a small representation of the supersymmetry algebra
(eight dimensional for a CPT conjugate multiplet):
\begin{equation}
\label{BPS}
m=\sqrt{2}\,\vert Z\vert .
\end{equation}
All the elementary excitations (quarks, W bosons, \dots ), as well as
the known solitonic states (monopoles and dyons) are BPS states
and their masses are thus given by (\ref{BPS}).
We will also obtain an interesting physical interpretation of
curious monodromy properties, and we will be able to compute
the monodromy matrices directly in the microscopic theory.
This study will finally solve some puzzles about the BPS mass formula
and the renormalization group flow.
\section{Presentation of the problem}
Along the Coulomb branch,
the central charge $Z$ of $N=2$ supersymmetric QCD contains, in
addition to the electric and magnetic charges $Q_e$ and $Q_m$, the 
``quark number'' charges $S_f$. These charges correspond to the invariance of
the lagrangian under the transformations $Q_f\mapsto
e^{i\alpha }Q_f$, $\tilde Q_f\mapsto e^{-i\alpha }\tilde Q_f$, where
$Q_f$ and $\tilde Q_f$ are the $N=1$ chiral superfields making up
a matter $N=2$ hypermultiplet. Classically $Z$ reads
\begin{equation}
Z_{cl}=2a\,\biggl({1\over g}\, Q_{e} + 
{i\over g}\, Q_{m}\biggr) +{1\over\sqrt{2}}\, 
\sum _{f=1}^{N_{f}}m_{f}S_{f},\label{Zcl}
\end{equation}
where $N_f$ is the number of flavours, $m_f$ the bare mass of the
hypermultiplet $(Q_f,\tilde Q_f)$, and $g$ the gauge coupling
constant. For the gauge group SU(2), on which I will
focus for conciseness,
the Dirac quantization condition can be written
$Q_m=4\pi n_m/g$, where $n_m$ is an integer also called the 
``magnetic charge.'' In \cite{SWII}, an exact quantum formula for $Z$ was
proposed: $Z=an_e+a_Dn_m+{1\over\sqrt{2}}\sum _f
m_fS_f$. Here $a$ is the Higgs expectation value, $\langle\phi
\rangle =a\sigma _3$, and $a_D={1\over 2}\partial _a {\cal F}(a)$ is 
the dual variable which can be expressed in terms of the prepotential
${\cal F}$ governing the low energy effective action. This formula
for $Z$ is a straightforward generalization of the corresponding
formula for the pure gauge theory derived in \cite{SWI} using
electric-magnetic duality arguments. 
However, we will see that interpreting $S_f$ in this formula as
being the physical quark number is not free from contradictions. Problems
also arise when considering that
the term
$a n_e+a_D n_m$ stems from corrections to the physical electric and
magnetic charges alone.

To understand the origin of these difficulties, let us set for a moment
the bare masses to zero.
In this case the quantum formula 
$Z=an_e+a_Dn_m$ is not obtained from (\ref{Zcl}) by simply
replacing $g$ by the running coupling constant, even
at one loop.
One also has to take into account that,
because CP invariance is spontaneously broken by 
$\IM a\not =0$, the physical electric charge can pick up
terms in addition to $gn_e/2$.
The simplest example of this phenomenon
was first studied by Witten in \cite{Witt}.
In the theories with zero bare masses,
all CP violation can be absorbed in a $\theta $ 
angle by performing a chiral 
$U(1)_R$ transformation.
The formula of
\cite{Witt} can then be readily adapted to our case and yields
$Q_e={g\over 2}(n_e-{4-N_f\over\pi }n_m \arg a)$. Note that
the $S_f$ charges are not affected by a $\theta $ term since the latter
appears in the lagrangian in front of $F\tilde F$ which does
not transform under $S_f$. Thus when $m_f=0$, $S_f$ is expected
to have the value one can compute in the CP conserving theories. For
instance, $S_f=\pm 1/2$ when $n_m=1$ \cite{JR}.

However, when the bare masses are non-zero, we would expect
$S_f$ to depend on the Higgs expectation value,
or alternatively on the gauge invariant
coordinate on the Coulomb branch, $u=\text{tr}\,\langle\phi ^2\rangle$.
This is strictly analogous to
the phenomenon 
first discovered in \cite{GW}, where irrational fermion number values
were found in some CP violating field theories.
At first sight, this seems bizarre. As $S_f$ are real numbers, 
a non-trivial $u$-dependence would seem to violate holomorphy. One
could then be tempted to forget about CP breaking and argue that,
because of supersymmetry, the $S_{f}$ charges must be constant 
and equal to the values one 
computes in CP conserving theories. But one then faces another,
more subtle, difficulty.
 Suppose one is studying the renormalization group flow, say from the
$N_{f}=1$ to the $N_{f}=0$ theory, and that in particular one is
trying to deduce the spectrum of stable BPS states of the $N_{f}=0$
theory from the one of the $N_{f}=1$ theory. What should occur is
that some states of the $N_{f}=1$ theory, becoming infinitely massive
in the process,
disappear from the spectrum of the $N_{f}=0$ theory, and that other
states, remaining of finite mass, finally constitute the stable BPS
states of the $N_{f}=0$ theory.
Since the work in \cite{FB}, we know that the spectrum of the
$N_{f}=0$ theory
is indeed strictly included in the spectrum of the $N_{f}=1$ theory.
Limiting the discussion to the weak-coupling spectra, which can be
described semiclassically, all the monopoles of odd $n_{e}$ disappear
when one goes from $N_{f}=1$ to $N_{f}=0$, while the monopoles of
even $n_{e}$ form the solitonic spectrum of the pure
gauge theory. But this is incompatible with the previous formula
for $Z$. If $S=\pm 1/2$ for the monopoles, their masses 
$m=\sqrt{2}\, |Z|$
will diverge
whatever their electric charge $n_{e}$, since $a_{D}$ and $a$ must flow
towards the solution of the pure gauge theory under the
action of the renormalization group.
So one must definitively give up the idea that the constants
$S_{f}$ appearing
in $Z$ could be the physical charges. I will rename these constants
$s_{f}$ and reserve the symbol $S_f$ for the physical charges.
The $s_f$ may be zero even for monopoles $n_{m}=1$.
The exact quantum formula for the central charge is then
\begin{equation}
\label{Zqua}
Z=an_e +a_Dn_m+{1\over\sqrt{2}}\,\sum _{f=1}^{N_f} m_f s_f.
\end{equation}
In the remainder of this letter I will explain where the
physical charges $S_f$ hide and how to compute the numbers
$s_f$. 
We will also find 
a physical explanation of the curious monodromy properties $a_{D}$
and $a$ have in the massive theories (they pick up constants in addition
to the standard $SL(2,\Z)$ transformations), and a new method to
compute the monodromy matrices, directly in the microscopic theory.
\section{The computation of the physical charges}
In this section, I outline the computation of the physical charges 
$S_f$ and of the electric charge, focusing on the contribution
of the fermions. This yields the most interesting results. A detailed
discussion of possible additional contributions, the
generalization to any gauge group, as well as the discussion
of other physical aspects related to CP invariance in our theories
shall be published elsewhere.

To study the semiclassical contributions to the $S_f$ charge
of the Dirac fermions $\chi _f$
belonging to the hypermultiplet $Q_f,\tilde Q_f$,
we need to quantize the Dirac field around a non-trivial
monopole background characterized by a fixed magnetic charge
$n_m$. The Dirac equation is
\begin{eqnarray}
\label{diraceq}
i\sqrt{2}\, \gamma ^\mu D_\mu\chi_f= 
&\bigl(&M+i\gamma ^5 N\bigr)\, \chi _f \nonumber \\
+{1\over\sqrt{2}}\,&\bigl(&(\RE m_f)\,\chi _f+ i(\IM m_f)\gamma ^5\, \chi _f
\bigr),
\end{eqnarray}
where $M=\RE\phi $ and $N=\IM\phi $. Usually one focuses on 
the (complex) zero
modes of this equation, whose number $k(n_m)$ is given by an index theorem
of Callias \cite{cal}. Each zero mode carries one unit of $S_f$ charge,
which shows that if $-S_{f,m}(n_m)$ is the minimal value $S_f$ can reach
in the monopole sector $n_m$, the set of allowed values of $S_f$
will be $\{-S_{f,m}(n_m),-S_{f,m}(n_m)+1,\ldots ,-S_{f,m}(n_m)
+k(n_m)\}$. If (\ref{diraceq}) were CP invariant, one would
deduce that $S_{f,m}(n_m)=k(n_m)/2$ because $S_f$ is odd under CP. 
However, in our case we cannot
forget about the massive modes of (\ref{diraceq}). 
When CP is violated, the density of
states having positive energy differs from the density of 
states having negative
energy. This means that the Dirac operator under consideration
has a non-zero Atiyah-Patodi-Singer $\eta $ invariant,\footnote{
Note that the operator is defined on an open space here, thus we are
dealing with a generalized $\eta $ invariant.} 
which formally reads $\eta =\sum _{E_n>0}1-\sum _{E_n<0}1$ where
$n$ labels the energy levels.
It is not difficult to relate the spectral asymmetry quantified
by $\eta $ to the $S_f$ charge, carefully taking into account
the fact that $S_f$ must be odd under CP. One finds
\begin{equation}
\label{eta}
S_{f,m}(n_m) = {1\over 2}\, \eta .
\end{equation}
Fortunately, the computation of the $\eta $ invariants of various
Dirac operator, and the 
application to charge fractionization, has been extensively studied
in the literature. These works were motivated not only for purely
field theoretic reasons, but also because of their important
phenomenological applications in the physics of linearly conjugated
polymers (see e.g\hbox{.} \cite{polym}). 
In particular, in \cite{NSI},
a very general method, 
applicable to our Dirac operator, was developped.
See also \cite{NSII}, and \cite{rapport} for a review. 
The result of the computation is
\begin{equation}
\label{S}
S_{f,m}(n_m)=-{n_m\over 2\pi}\, \arg {a + m_f/\sqrt{2}\over
m_f/\sqrt{2} -a}\cdotp
\end{equation}
The same type of technique can be applied to evaluate the physical
electric charge, which picks up terms in addition to the standard
Witten effect \cite{Witt} term. One can for instance use the Gauss
law, following \cite{NSIII}, and find for our theory
\begin{equation}
\label{Q}
{2\over g}\, Q_e = n_e-{4\over\pi }\, n_m \arg a
+{n_m\over 2\pi}\,\sum _{f=1}^{N_f}\arg (m_f^2-2a^2).
\end{equation}
\section{Physical analysis}
Let us discuss the physical meaning of the formulas 
found in the previous Section. 
Note that we have a singular point when $a=\pm m_f/\sqrt{2}$,
which corresponds to a quark becoming massless.
It is very instructive
to study the monodromy properties of $S_f$ around this singularity.
Since all the non-trivial, $u$-dependent part of $S_f$ is
included in $-S_{f,m}$, encircling the singularity
at $a=m_f/\sqrt{2}$ yields $S_f(u)\mapsto S_f(u)-n_m$. This is
reminiscent of the shift $s_f\mapsto s_f+n_m$ the
constant $s_f$ undergoes \cite{SWII}.\footnote{Though constants, the
$s_f$ do transform non trivially when encircling a singularity,
in the same sense as the constants $n_e$ and $n_m$ are mixed by the
monodromy matrix.}
However, the sign difference between the transformations of $S_f$ and
$s_f$ is crucial. 
It definitively proves that $s_f$ cannot be identified with
$S_f$. Moreover, it shows that $S_f(u)$ and $a_D(u)$ pick up the
same term under the monodromy. This simply means that $S_f(u)$
is already included in $a_D(u)$ (at weak coupling), and is responsible
for the curious constant shift $a_D$ was known to undergo
since the work in \cite{SWII}.
In the strong coupling
region, since $a$ and $a_D$ are intimately related due to the
non-abelian monodromies, the
$S_f$ charges will also contribute to $a$. Note however that the
distinction between $Q_e$, $Q_m$ and $S_f$ is very unclear in
the strong coupling region, and that the natural quantities
to use are $a$ and $a_D$.

The fact that the variables $a$ and $a_D$ do pick up 
contributions from the electric, magnetic {\em and} $S_f$ charges
is very interesting from the physical point of view.
This means that duality 
in the theories with non zero bare masses 
is really an electric-magnetic-$S_f$ duality! This phenomenon
is likely to be quite general when abelian charges appear in the
central charge in addition to the electric and magnetic charge.

Now, it should be clear that the variables $s_f$ appearing
in the formula (\ref{Zqua}) are just constant parts of the
$S_f$ charges not already included in $a_D$ and $a$.
In particular one can compute $s_f$ in the weak coupling region
by studying the asymptotics of $a_D$ when $u$ and $m_f$
are large comparing to the dynamically generated scale of the theory,
then extract the terms contributing to $S_f$ from this asymptotics,
and choose $s_f$ in order to match with the formula
$S_f=-S_{f,m}(n_f)+p$. Here $p$ is an integer between $0$ and
$k(n_f)$, see Section 2.

Let us close this Section computing
the $SL(2,\Z )$ monodromy matrix $M$
corresponding to a singularity due to a quark becoming massless.
This can easily be done using (\ref{Q}) which shows that
$Q_e$ contribute to $a_D$  through the term
$-{4\over\pi}a\arg a+{1\over 2\pi}a\,\sum\arg (m_f^2-2a^2)$,
see (\ref{Zcl}).
Thus, when no bare masses coincide,
\begin{equation}
M=\pmatrix{1 & 1\cr 0 & 1\cr }
\end{equation}
since $a$, being a good local coordinate around $u=m_f^2$,
obviously does not transform. This result agrees with the 
standard computation from the low energy effective theory. 
When some bare masses coincide, we obtain the monodromy
matrix corresponding to several hypermultiplets becoming massless at the
same time. Then,
performing $SL(2,\Z)$ transformations, it is possible to deduce
the most general monodromy matrix corresponding to any number of $(n_m,n_e)$
states becoming massless. 
\section{conclusions}
We gained interesting physical insight 
in the meaning of duality in $N=2$ supersymmetric QCD by using
semiclassical methods. This was possible since when bare masses are
much larger than the dynamically generated scale of the theory,
some singularities can be present at weak coupling. We found that
in this regime
the contribution of the physical electric charge $Q_e$ to $a_D$ 
can account for the $SL(2,\Z)$ monodromy matrices associated
with the weak coupling singularities.
This provides a new way to derive these matrices, directly in the
microscopic theory.
More important, we saw that the physical
$S_f$ charges contribute to $a_D$ and
can account for the constant shift this variable undergoes.
At strong coupling, the {\em three} abelian charges appearing
in the central charge of the supersymmetry algebra will
intimately mix together, providing an example of an
electric-magnetic-quark number duality. 
\section*{Note added}
After the first appearance of this work on the hep-th archive (9609101), 
a preprint appeared \cite{CERN} where the scenario described in
Section 1 for the renormalization group flow is shown to occur.
For the $N_f=1$ theory, the authors of 
\cite{CERN} were able to find, within a string theory
framework, that $s=0$ for the monopole $(1,0)$ and $s=-1$ for the
dyon $(1,1)$. This would contradict the semiclassical computation
if $s$ were the physical charge, but is in perfect agreement 
with the discussion of Section 1.
\section*{Acknowledgments}
I wish to thank Adel Bilal, Eug\`ene Cremmer and Jean-Loup Gervais
for useful discussions and encouragements, as well as
Tom Wynter who kindly read the manuscript.

\end{multicols}
\end{document}